# Semiclassical planetology: some results

Vladan Čelebonović
Institute of Physics,Pregrevica 118,11080 Zemun-Beograd,Yugoslavia
vladan@phy.bg.ac.yu
vcelebonovic@sezampro.yu

Abstract: In the early sixties two Yugoslav scientists, Pavle Savić and Radivoje Kašanin have started developing a theory of the behaviour of materials under high pressure. The initial work was published between 1962 and 1965. In the course of time, this theory has found applications in planetology and laboratory high pressure work. The aim of this lecture is to review the basic physical ideas and some results of the planetological applications of this theory.

## Introduction

Planetology is defined as a branch of astronomy and physics pertaining to the study of planets. Its origins can not be traced precisely, because first purely logical and philosophical considerations on the "wandering stars" were made already in antiquity. First scientific studies of the problem of the origin of planets date to the times of Kant and Laplace . However, studies of planetary interiors had to wait until the second half of the last century and the development of seismology. It has been shown on several occasions in the XIX century that seismic waves , originating in earthquakes, in their propagation through the interior of the Earth encounter regions of different densities [1]. These were the first experimental proofs that the Earth's interior is stratified.

The turn of the century witnessed first attempts of laboratory studies of planetologically important materials. Water, for example, has been subdued to increased pressure, already in XVII century. It is known that an Englishman, a certain Mr.Cannon , compressed water to 0.01 GPa at room temperature, and obtained water ice [2] . This attempt is nowadays remembered just as a detail of historical interest . Laboratory experiments on the behaviour of water under high pressure started due to work by Tamman [3] and Bridgman [4] in the first decade of the XX century . For comparison, note that there exist at least 15 phases of ice, and that water has been studied in diamond anvil cells at values of pressure $P \leq 128 GPa$ [5] .

Theoretical planetology started to develop rapidly in this century. The usual process of modellization of planetary internal structure contains the following steps: assume some chemical composition of the body under study and some form of the equation of state of the material which makes up the planet, and then calculate the radial distribution of pressure, density and temperature ( see, for example, [6] ) .

The aim of this lecture is to review the main physical ideas of a theory of behaviour of materials under high pressure, proposed in the sixties by P.Savić and R.Kašanin (nicknamed the SK theory) [7] . As the theory has applications in astrophysics, as well as in laboratory high pressure studies, we shall here present only a part of the planetological results . For a previous review of the theory see [7a].

The SK theory

This theory started developing after a paper by Savić which had the aim to explore the origin of planetary rotation [8] . It was shown in that paper that in order to draw conclusions about the origin of planetary rotation, one needed to know details about their internal structure. The mean planetary densities were fitted by a simple analytical formula

$$\rho = \frac{4}{3} 2^{\varphi} \qquad (1)$$

Note that the factor $\frac{4}{3}$ turns out to be very close to the mean solar density, and $\varphi$ is an integer: $\varphi \in \{-1,0,2\}$. Choosing values of $\varphi$ by trial and error, it becomes possible to reproduce the known values of the mean planetary densities. The original data presented in [8] are reproduced in Table I:

Table I

| Object | $\varphi$ | $\rho_{calc}$ $(kg/m^3)$ | $\rho_{obs}$ $(kg/m^3)$ |
|---|---|---|---|
| Saturn | -1 | 660 | 650 |
| Sun | 0 | 1330 | 1410 |
| Jupiter | 0 | 1330 | 1340 |
| Uranus | 0 | 1330 | 1360 |
| Neptune | 0 | 1330 | 1320 |
| Earth | 2 | 5320 | 5520 |
| Mars | 2 | 5320 | 3940 |
| Venus | 2 | 5320 | 5210 |
| Mercury | 2 | 5320 | 5600 |

The symbols $\rho_{calc}$ and $\rho_{obs}$ denote, respectively, the density calculated according to eq.(1) and values derived from known planetary masses and radii.

The importance of eq.(1) is in the fact that it helped the authors of SK to introduce a new idea in planetology - that the internal structure of a planet is influenced to a large extent by changes , provoked by high pressure, of the atomic and/or molecular structure of the mixture of materials that the planet is made of, as well by changes in the lattice structure of these materials. Apart from the "quantum " nature of eq.(1), they were pushed in that direction by two more factors: the geophysical data on the propagation of seismic waves through the Earth, which showed that the Earth's interior consisted of different layers with sharp transitions between them, and laboratory experiments on materials under high pressure (such as [9] ).

At the time, they were not aware of the fact that Enrico Fermi invoked the possibility of changes of atomic and molecular structure under high external pressure several decades before them, while testing the applicability of the then "new" Schrödinger equation. However, Savić and Kašanin were the first to use the idea of changes of atomic structure under high pressure in planetology. This phenomenon was termed "pressure excitation",and received full quantum mechanical treatement only recently [10].

In four years following the publication of [8] Savić and Kašanin developed their theory of the behaviour of materials under high pressure [7] . It has two general aims: to develop a theoretical explanation of the origin of rotation of cold solid celestial bodies ( such as planets,satellites,asteroids) and a theory of behaviour of dense matter. The result is a theoretical description of some aspects of phenomena occuring in materials under high pressure, founded on known laws of classical physics combined with some facts from quantum mechanics - hence the name semiclassical .

The considerations of SK start from a low temperature isolated cloud of gas and dust, consisting of an arbitrary number of chemical elements and their compounds. The evolution in time of such a cloud is governed by two physical processes: the mutual interactions of its constituting particles, and the loss of energy due to radiation. Because of the low temperature of the cloud, this radiation is somewhere in the red or even IR part of the spectrum .

The obvious consequence of these two factors is the increase of density, and pressure in the interior of the cloud. A further consequence is the excitation and ionisation of the atoms and molecules in the cloud. Pressure excitation and ionisation is an expectable consequence of the perturbation of the electronic energy levels by the external pressure field. The physical possibility of such a coupling can be proved even in solvable elementary quantum mechanical systems such as a finite potential well. Increasing the pressure to which a system is subdued leads to the expansion of the radial part of the electronic wave functions of the atoms and molecules that make up the system.

When the pressure has risen sufficiently so that pressure ionisation can start, a phase transition occurs in the cloud - it passes into the state of a poly-component plasma. This plasma consists of a randomly moving electron gas and neutral and ionized atoms and molecules. Such an electron gas has a non-zero magnetic field [11]. Owing to a combination of high pressure and low temperature, the magnetic moments of the ionized atoms and molecules become oriented in paralel, and the resulting torque starts the rotation of the whole system. We have here given a qualitative description of the process which, according to SK, gives rise to rotation. Details, pertaining to the rotation of the Moon and the form in which the speed of rotation is expressed in SK are given in [7],[12].

The second line of investigations pursued by SK is the study of the behaviour of materials under high pressure. Experimental work in this domain is linked with a multitude of problems (such as filling a diamond anvil cell, or fixing the contacts on a specimen). On the theoretical side, a specimen of any material is a typical example of a many-body system, whose Hamiltonian has the form:

$$H = \sum_{i=1}^{N} -\frac{\hbar^2}{2m}\nabla_i^2 + \sum_{i=1}^{N} V(|\vec{x}_i|) + \sum_{i,j=1}^{N} v(|\vec{x}_i - \vec{x}_j|) \qquad (2)$$

All the symbols in eq.(2) have their standard meanings. Instead of embarking on a calculation of the thermodynamical functions starting from eq.(2) Savić and Kašanin have based their theory on a set of six experimentally founded premisses supplemented by a selection rule.

The mean interparticle distance $a$ is defined by the relation

$$N_A(2a)^3 \rho = A \qquad (3)$$

where $N_A$ denotes Avogadro's number, A the mean atomic mass of the material and $\rho$ is the mass density. As a next step, one introduces the "accumulated" energy per electron as

$$E_a = \frac{e^2}{a} \qquad (4)$$

It can be shown [13] that the mean interparticle distance defined in eq.(3) is a multiple of the Wigner-Seitz radius which contains the ionic charge Z.

The basic premisses of the SK theory are [7],[12] :

1. The density of a material is an increasing function of the applied pressure.
2. With increasing density, a material undergoes a sequence of first order phase transitions. Phases are indexed by an integer $i$. The phase ending at the critical point is denoted as the zeroth phase. For an arbitrary phase, the density is delimited by :

$$\rho_i^0 \leq \rho_i \leq \rho_i^*$$

or  $$\frac{\rho_i^*}{\alpha_i} \leq \rho_i \leq \rho_i^* \qquad \alpha_i \geq 1 \qquad (5)$$

3. Assuming in eq.(1) that $\varphi_{i+1} - \varphi_i = 1$ one gets that the maximal densities in two sucessive phases are related by

$$\rho_{i+1}^* = 2\rho_i^* \qquad (6)$$

4. It is assumed that

$$\frac{E_i^*}{E_i^0} = \frac{E_{i+1}^0}{E_i^*} \qquad (7)$$

A relationship between the accumulated energies in sucessive phases was needed so as to render the calculations possible, and eq.(7) was assumed because of its simplicity. It can be shown that $\alpha_i \alpha_{i+1} = 2$ and that

$$\alpha_i = \begin{cases} 6/5, i=1,3,5,... \\ 5/3, i=2,4,6,... \end{cases} \qquad (8)$$

5. The final density of the phase $i = 0$ is given by

$$\rho_0^* = \frac{A}{3\overline{V}} \qquad (9)$$

where $\overline{V}$ is the molar volume of the material under standard conditions [12].

6. It follows from assumption 3 that

$$\frac{1}{\bar{\rho}} = \frac{1}{2}\left(\frac{1}{\rho_2^0} + \frac{1}{\rho_2^*}\right) \quad (10)$$

where the symbol $\bar{\rho} = \frac{A}{\bar{V}}$ is the density of a material at the zero point.

These assumptions enable the calculation of the value of pressure at which a first order phase transition can be expected in a material. Mathematically, the idea of this calculation is to compare the work performed by the external pressure on compressing the material with the corresponding change of the accumulated energy. Details concerning this algorithm, as well as its applications to 20 materials chosen at random and an analysis of the possible causes of the discrepancies can be found in the literature [12] - [14].

The algorithm proposed in the SK theory gives the mathematical rule for calculating the phase transition pressure. However, SK have devised a selection rule for determining those phase transitions which are physically possible [7]. A transition $i \rightarrow i+1$ is possible if

$$E_0^* + E_I = E_i^* \quad (11)$$

where $E_I$ denotes the ionisation and/or excitation potential.

Applications in planetology

In planetology, the SK theory gives the possibility of modelling the internal structure of planets, satellites and asteroids. It can not be applied to stellar structure studies simply because it does not take into account the fact that nuclear reactions take place in stars.

The input data needed for a planetological application of this theory are the mass and the radius of the object under study. SK have developed algorithms which, starting from this pair of values, allow the determination of the number and thickness of layer which exist in the interior of the object, the distribution of pressure density and temperature with depth, the strength of the magnetic field and the allowed interval in which the angular speed of rotation of the object can be expected. This theory also gives the mean atomic mass of the chemical mixture that the object under study contains.

The SK theory has been applied to the Sun and all the planets except Saturn and Pluto, the Moon, the Galileian satellites , the 5 big satellites of Uranus and the asteroids 1 Ceres and 10 Hygiea. It was known to SK that the model of the solar interior within their theory can not be realistic. Interestingly, the central temperature obtained is of the correct order ( $T_c \approx 10^7 K$ ).

The form of the results is illustrated in the following tables, which contain the models of the Earth and Moon calculated according to the SK theory. The maximal values of the temperature in various layers were calculated according to [15]

## Table II

### the interior of the Earth

| depth (km) | 0-39 | 39-2900 | 2900-4980 | 4980-6371 |
|---|---|---|---|---|
| $\rho_{max} (kg/m^3)$ | 3000 | 6000 | 12000 | 19740 |
| $P_{max} (Mbar)$ | 0.25 | 1.29 | 2.89 | 3.7 |
| $T_{max} (K)$ | 1300 | 2700 | 4100 | 7000 |

<A> = 26.56

## Table III

### the interior of the Moon

| Depth (km) | 0-338 | 338-1738 |
|---|---|---|
| $\rho_{max} (kg/m^3)$ | 3320 | 6640 |
| $P_{max} (Mbar)$ | 0.015 | 0.089 |
| $T_{max} (K)$ | 529 | 793 |

<A> = 71

Parameters of the models of the Earth and the Moon, and similar models of other objects calculated within the SK theory, can not be compared to in situ experimental data, but only to observable consequences on the surfaces of these objects of conditions in their interiors. However, the mean atomic mass <A> and combinations of chemical elements and/or compounds which fit it can be compared to experimental data obtained by remote spectroscopy from the Earth and/or space probes. A table of values of <A> is avaliable in [12] , and a distribution of <A> with radial distance from the Sun is represented in Fig.1

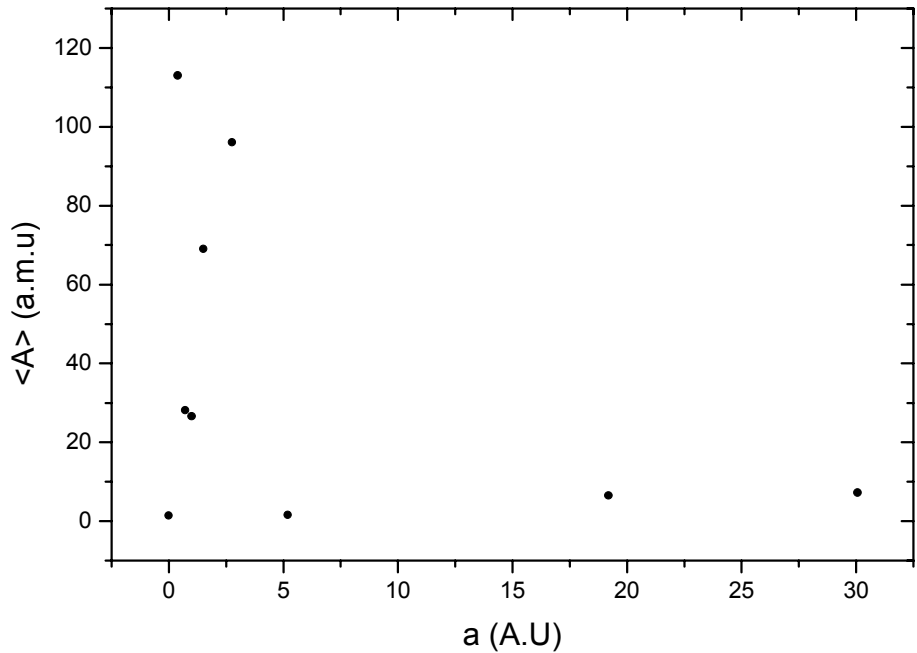

Figure 1 : the distribution of <A> with radial distance from the Sun

The radial distribution of <A> is of considerable interest for cosmogony, because it is a consequence of compositional gradients and transport processes which existed and operated in the accretion disk from which the Solar System condensed. Namely, values represented in Fig.1 were derived within the SK theory, whose predictions have been compared with laboratory results [13] and astronomical observation. On the other hand, direct experimental verification of cosmogonical calculations is nearly impossible.

The most sucessful example of the determination of the value of <A> for an object is the case of the Galileian satellites. Their values of <A> were determined in [16] back in 1987. First in-situ data on their chemical composition were obtained by the "Galileo" space probe in 1996. It turned out that predictions made by the SK theory in 1987., are in excellent agreement with experimental data collected nearly 10 years afterwards [17] . At the time of publication, [17] has arisen considerable interest.

Another interesting case of the determination of <A> has been the asteroid 10 Hygiea [18]. It is known from observation that asteroids 1 Ceres and 10 Hygiea are spectroscopically similar.Using the known values of the mass and radius of Ceres, its value of <A> was calculated. Combining this result with the value of R for 10 Hygiea as determined by IRAS, it became possible to calculate the mass of this asteroid.

Conclusions

In this lecture we have reviewed some planetological applications of a semiclassical theory of dense matter proposed by Savić and Kašanin in the sixties, and later applied and developed by various authors. Naturally, no details of the calculations were given on the text, but they can be found in the references .

The SK theory is mathematically simple. It has been developed with the aim of describing extremely diverse and complicated phenomena, but this complexity has been bypased by a set of basic postulates and a selection rule.This is a great advantage of this theory over better known models of phase transitions . However, there is also a hidden disadvantage - namely, the simplification of the physical assumptions necessarily renders a calculation cruder, and increases the discrepancies between the calculated and experimental results.

We have discussed in this contribution some of the planetological results of the SK theory, and the obvious question is " What lies ahead?" . This theory has a wide field of possibilities for future improvements and applications .

For instance, it would be very useful to improve its basic postulates. Eq.(1) should be re-derived from modern data, and it is very probable that a new set of values of the exponent $\varphi$ would emerge from this.The point here is that this equation is later used in calculation of the phase transition pressure, because it gives the ratio of densities in two sucessive phases. In the same domain, it would be important to develop a microscopic explanation of eq.(1) and,in particular of the physical meaning of $\varphi$ . On the laboratory side, it would be useful to repeat the analysis presented in [13], but for a larger set of materials. It would be interesting to try to include nuclear reactions in the theory and thus render it applicable to stellar structure studies.

On the planetological side, there remain two planets which have not been modelled: Saturn and Pluto. But more interesting than that could be modellization of the asteroids. Using similarities, as it has been done for Hygiea and Ceres, could be a useful method for determining masses of some asteroids.

In cosmogony,the SK theory could be also be useful. Just one example is its possibility to determine the value of <A> .This implies that one can, starting from observed data, gain knowledge about the distribution of chemical elements within the planetary system, and thus constrain cosmogonical models. Such studies have already been performed on the Jovian and Uranian satellite systems [16],[19]. It has also been shown that Triton and Neptune are widely different by their chemical composition [20] , which is in perfect agreement with the result known in celestial mechanics that Triton is a captured body.

To conclude , we can say that this simple theory of complicated phenomena has achieved interesting planetological and laboratory results and that it offers possibilites for future work both in astronomy and physics.